\documentclass[newstyle,twocolumn,proceedings]{rmaa}
\usepackage{rmaacite}

\newcommand{\Macc}{$\dot M_{\rm acc}$}
\newcommand{\Mj}{$\dot M_{\rm j}$}
\newcommand{\msunyr}{$M_{\odot}$ yr$^{-1}$}
\newcommand{\fprot}{$f_{\rm p}$}
\newcommand{\fion}{$f_{\rm i}$}
\newcommand{\fe}{$f_{\rm e}$}
\newcommand{\nel}{$n_{\rm e}$}
\newcommand{\oi}{[O~{\sc i}]}
\newcommand{\sii}{[S~{\sc ii}]}
\newcommand{\nii}{[N~{\sc ii}]}

\title{Atomic T Tauri disk winds heated by ambipolar diffusion}
\author{
        Paulo Garcia
        \affil{Centro de Astrof\'{\i}sica da UP, Portugal \&  CRAL, UMR 5574 du CNRS, France (pgarcia@astro.up.pt)}
\and    Sylvie Cabrit
        \affil{Observatoire de Paris, DEMIRM, UMR 8540 du CNRS
          (sylvie.cabrit@obspm.fr)}
\and    Jonathan Ferreira and Fabien Casse
        \affil{Laboratoire d'Astrophysique, Observatoire de Grenoble,
        CNRS/UJF UMR 5571 (ferreira@obs.ujf-grenoble.fr)}
\and    Luc Binette
        \affil{Instituto de Astronom\'{\i}a, UNAM (binette@astroscu.unam.mx)}
}

\fulladdresses{\item   P.  Garcia,   Centro  de   Astrof\'{\i}sica  da
Univ. Porto, Rua das Estrelas s/n, 4150-762 Porto, Portugal
                \item S. Cabrit, DEMIRM, Observatoire de Paris, 
61 Avenue de l'Observatoire, F-75014 Paris
                \item J. Ferreira and F. Casse, LAOG,
Observatoire de Grenoble, BP 53, F-38041 Grenoble Cedex 9
                \item L. Binette, Instituto de Astronom\'\i a,
 UNAM, Ap. 70-264, 04510 D. F., M\'exico

}

\shortauthor{Garcia et al.}
\shorttitle{Atomic T Tauri disk winds heated by ambipolar diffusion}

\keywords{magnetohydrodynamics --- ISM: jets and outflows --- Stars:
  pre-main-sequence}

\abstract{We summarize results on the thermal and ionization
structure of self-similar, magnetically-driven, atomic disk winds heated by
ambipolar diffusion. We improve upon earlier work by Safier by considering
(1) new MHD solutions consistent with underlying cold keplerian disk
equilibrium, (2) a more accurate treatment of the micro-physics, and (3)
predictions for spatially resolved forbidden line emission (maps, long-slit
spectra). The temperature plateau $\simeq 10^4$~K found earlier is
recovered, but ionization fractions are revised downward by a factor of 10,
due to previous omission of thermal speeds in ion-neutral momentum-exchange
rates. The physical origin of the temperature plateau is outlined.
Predictions are then compared with T~Tauri star observations, with emphasis
on the necessity of suitable beam convolution.  Jet widths and variations
in line profiles with distance and line tracer are well
reproduced. However, predicted maximum velocities are too high, total
densities too low, and the low-velocity \oi\ component is too weak. Denser,
slower MHD winds from warm disks might resolve these discrepancies.}

\listofauthors{Garcia, P., Cabrit, S., Ferreira, J., Binette, L., Casse, F.}
\indexauthor{}
\indexauthor{}

\begin{document}

\maketitle

\section{Introduction}
\label{sec:intro}

Collimated mass ejection in young T~Tauri stars (TTS) is intimately
correlated with the accretion process (Cohen et al. 1989; Cabrit et al.
1990; Hartigan et al.  1995; hereafter HEG95).  It is currently believed
that magnetic forces are responsible for both the high ejection efficiency
(\Mj/\Macc $\simeq 0.01-0.1$; Hartigan et al. 1994, HEG95) and the high
degree of collimation of these winds.  Yet, the exact field structure and
flow dynamics are not well established. As a result, the role played by
magnetized ejection in the physics of TTS, in particular in extracting
angular momentum from the circumstellar disk and/or stellar surface,
remains a major enigma.

Ejection from a wide range of disk radii has been invoked to explain the
compact low-velocity \oi\ emission peak observed in accreting TTS (Kwan \&
Tademaru 1988, 1995; Hirth et~al.  1994; HEG95; Kwan 1997).  However, it is
still unclear whether the high-velocity ``microjets'' observed towards
several TTS could trace the outer collimated regions of the same disk wind,
or whether they require a distinct ejection component.

A new tool to discriminate between models consists in confronting
theoretical predictions with recent {\it spatially resolved} observations
of the inner wind structure of TTS in optically thin forbidden lines of
\oi, \sii, and \nii\ (see Bacciotti et al., Dougados et al., Ray et al.,
this volume, and refs. therein).

Of the wide variety of models available in the literature, only two classes
have self-consistent stationary MHD solutions that have been used
for detailed observational predictions: (1) {\it `disk winds'} from a wide
range of disk radii (Blandford \& Payne 1982; Ferreira 1997) and (2)
{\it `X-winds'} from a tiny region of open stellar field lines near the
disk corotation radius (Shu et~al. 1995). Both are in the regime of ``cold''
MHD, where ejection is only possible by magneto-centrifugal launching on
field lines sufficiently inclined from the disk axis, as thermal energy
is insufficient to cross the potential barrier (unlike in the solar wind).

Synthetic maps and long-slit spectra were first presented by Shang et
al. (1998) and Cabrit et al. (1999) for X-winds and disk winds
respectively, using a parametrized temperature and ionization fraction.
More reliable predictions require actually solving for the wind thermal and
ionization state, given some local heating mechanism. Shock heating was
recently invoked to explain line ratios in the outer regions of microjets
(Dougados et al., Lavalley-Fouquet et al. 2000) but this mechanism is not
fully consistent with a stationary solution and involves additional free
parameters.  Other modes of mechanical energy dissipation (e.g. turbulence)
have the same drawbacks.

Here, we consider only a heating process intrinsic to MHD winds of low
ionization, namely ion-neutral frictional drag ({\it `ambipolar diffusion
heating'}), which was first applied to predict integrated forbidden line
emission in disk winds by Safier (1993a,b). It requires no extra free
parameter and is self-consistently determined by the disk wind MHD
structure. It will yield the ``minimum'' possible emission fluxes and
ionization level, but emission maps and line profiles should still provide
useful tests of the model collimation and dynamics. A detailed account of
our results is given in Garcia et al. (2001).

\section{Thermal and ionization structure}
\label{sec:thermal}

\subsection{Method and improvements over previous work}

We follow the  same general approach as in the pioneering work by Safier
(1993a).  We  follow a  fluid  cell  along  a given  flow  streamline,
characterized by  $\varpi_0$, the field  line footpoint radius  in the
disk midplane,  and solve for  its coupled temperature  and ionization
evolution, including all relevant heating and cooling terms.  We check
a posteriori that the single, cold fluid assumptions made in computing
the  dynamical MHD  solution  remain valid,  i.e.   that drift  speeds
between ions and  neutrals are small compared to  the bulk flow speed,
and that  thermal pressure gradients are negligible  compared to other
forces.   Our equations,  heating/cooling terms,  and self-consistency
checks are described in Garcia  et al.  (2001). Important
improvements to the work of Safier (1993a) include:

- Use of new self-similar cold  disk winds  solutions  from Ferreira
(1997), where vertical equilibrium  of the keplerian accretion disk is
treated self-consistently. The mass loading and magnetic lever arm are
then fixed by a single  parameter: the ejection efficiency in the disk
$\xi = d \ln \dot{M}_{\rm acc}/d \ln \varpi_0$.

- Treatment of the ionization evolution of heavy elements (C, N, O, S,
Ca, Mg, Fe, \dots),  photo-ionization heating, and cooling by hydrogen
recombination lines, using the MAPPINGS~Ic code (Binette et al., 1985,
Binette \& Robinson 1987).

- Correct  computation  of  ion-neutral  momentum exchange  rates  for
ambipolar diffusion, including thermal speeds (cf.~Draine 1980).

- Depletion of heavy atoms into grains outside
the dust sublimation cavity \cite{Savage96}.

\subsection{MHD solution parameters}

To limit the  number of free parameters, we fix  the disk aspect ratio
$\varepsilon  = h/r  = 0.1$,  as  indicated by  observations of  HH~30
(Burrows  et  al.   1996),  and  set  the  magnetic  turbulence  level
parameter   (controlling  disk  accretion   across  field   lines)  to
$\alpha_m=1$ (cf. Ferreira  1997).  The central star mass  is taken as
0.5~$M_\odot$.  Thus only 2 free global parameters are left:

- The  ejection  efficiency parameter  $\xi$,  which typically  ranges
between 0.005  and 0.01 for the above  $\varepsilon$ and $\alpha_{m}$.
Our disk wind extends from $\varpi_i$ to $10 \times \varpi_i$ and thus
$\xi \sim \dot{M}_{\rm jet}/\dot{M}_{\rm acc}$ as observed in TTS.

- The  disk  accretion rate,  \Macc,  which  determines  both the  jet
density scalings and the photoionizing UV flux from the accretion disk
boundary  layer. We  consider values  between $10^{-8}$  and $10^{-5}$
\msunyr,  covering the observed  range in accreting TTS  (HEG95).

\subsection{Results}

\begin{figure}[t]
  \includegraphics[width=\columnwidth]{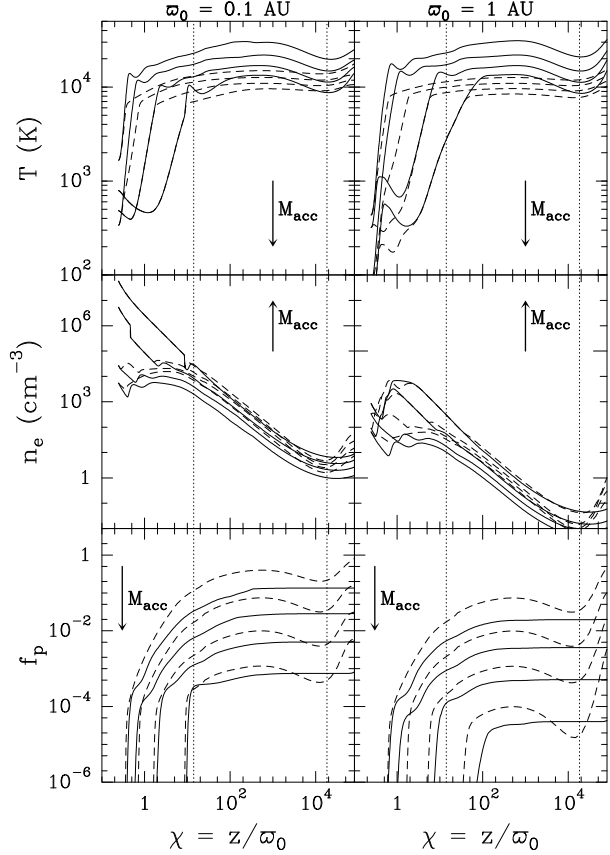}
  \caption{Temperature $T$, electronic density \nel, and proton fraction
\fprot =  $n(H^+)/n_H$ along two streamlines anchored  at $\varpi_0$ =
0.1~AU (left column) and $\varpi_0$  = 1~AU (right column), for a disk
wind  with   $\xi  =$  0.007.  For  comparison,   dashed  curves  show
calculations  with  local   ionization  equilibrium  imposed.   \Macc\
increases in the  direction of the arrow from  $10^{-8}$ and $10^{-5}$
\msunyr\ by factors of 10. Dotted vertical lines mark the locus of the
Alfv\'en  surface  ($\chi  \simeq   10$)  and  the  point  of  maximum
streamline  opening ($\chi \simeq  2 \times  10^4$) before refocusing}
\label{fig:results}
\end{figure}

\begin{figure}\label{fig:heatcool}
  \includegraphics[width=\columnwidth]{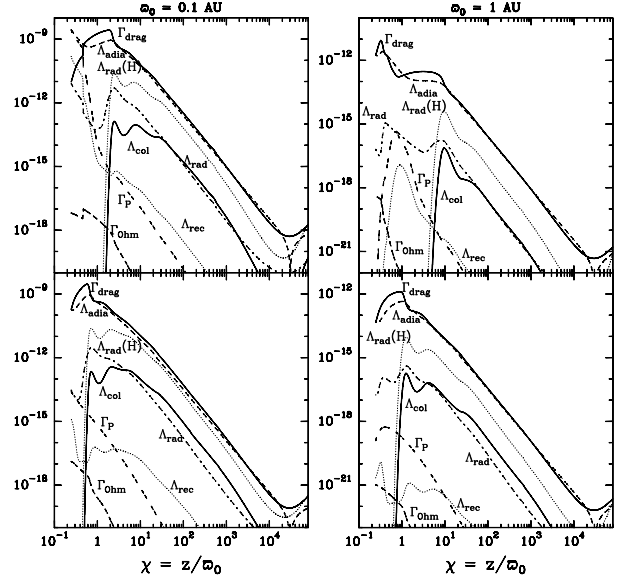} 
\caption{Heating
and cooling terms (in erg s$^{-1}$ cm$^{-3}$) for the same model as in
Fig.~1.    Top:    \Macc=$10^{-6}   M_\odot   \rm{yr}^{-1}$,   Bottom:
\Macc=$10^{-7} M_\odot \rm{yr}^{-1}$.}
\end{figure}

Figure \ref{fig:results} presents the computed temperature, electronic
density \nel, and proton fraction  \fprot\ along two streamlines for a
typical solution with an intermediate $\xi$ value of 0.007 and various
mass accretion rates. The main results are the following:

- Temperature reaches a  plateau around $10^4$~K over most  of the jet
extent for the range of parameters applicable to T Tauri disk winds.

- The proton  fraction \fprot\  rises steeply at  the wind  base, then
freezes out in the far jet  region ($\chi = z/\varpi_0 \ge 300$) where
densities   are  low  and   ionization  timescales   exceed  dynamical
timescales.  In the temperature  plateau, \fprot\ is roughly inversely
proportional to \Macc\ and $\varpi_0$.

- The electron density \nel\  is dominated by photoionization of heavy
elements at the  wind base, but is weakly  dependent on \Macc\ further
out.

- The  main  heating  process  is  ambipolar  diffusion,  $\Gamma_{\rm
drag}$,   and  the   main   cooling  term   is  adiabatic   expansion,
$\Lambda_{\rm  adia}$.  A  close   match  between  the  two  terms  is
established    in     the    temperature    plateau     region    (see
Fig.~\ref{fig:heatcool}).

Remarkably,  the   same  behaviors   were  found  for   the  solutions
investigated by  Safier (1993a).  In the next  section we  explain why
they are generic properties of self-similar cold MHD disk winds.

\subsection{Physical origin of the hot temperature plateau}

%fig4_____________________________________________________________________ 
\begin{figure*}
\begin{center}
\resizebox{2\columnwidth}{!}{\rotatebox{-90}{\includegraphics{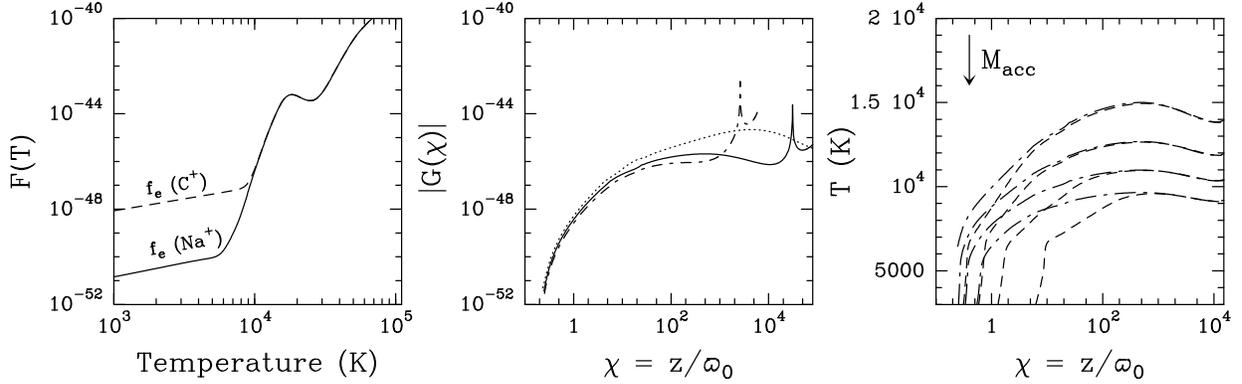}}}
\caption{ {\bf Left:} $F(T)$ in erg~g~cm$^3$~s$^{-1}$
{\em assuming local ionization equilibrium} and 
 photoionization of Na only (solid) or C (dashed). {\bf
Center:} $G(\chi)$ in  erg~g~cm$^3$~s$^{-1}$ for models A, B,
C  (bottom to  top), $\varpi_0=0.1$ AU  and 
\Macc\ $=10^{-6}$ \msunyr. {\bf  Right:} Temperature for model B assuming
ionization equilibrium (dashed), compared with $T_{\rm c}(\chi)$ as given
by Eq.~\ref{eq:Tc} (dot-dashed).  $\varpi_0=0.1$ AU and \Macc\ = $10^{-8}$
to $10^{-5} {M}_{\odot} \rm{yr}^{-1}$ , from top to bottom.}
\label{fig:plateau}%
\end{center}
\end{figure*}
%_____________________________________________________________________ 
%

It is useful to write the temperature evolution equation as a function
of $\chi = z/\varpi_0$ as:
\begin{eqnarray}
\frac{d  \ln  T}{d \ln \chi} &=&
- \frac{2}{3}  \frac{ d  \ln  \tilde{n}}{ d \ln \chi} \times
\Big(\frac{\Gamma_{\rm drag}}{\Lambda_{\rm adia}}-1\Big) \nonumber \\
&=& \delta^{-1} \times
\Big(\frac{G(\chi)}{F(T)}-1\Big), \label{eq:amb_temp}
\end{eqnarray}
where $\tilde{n}$ is the wind particle density, $\delta$ is a positive
function of order 1 before jet recollimation,
\begin{eqnarray}
G(\chi)=-\frac{\frac{1}{c^2} \bigl\| \vec{J} \times \vec{B}  \bigr\|^2}
{\tilde{n}^2 (\vec{v}\cdot\vec{\nabla})\tilde{n}} 
\end{eqnarray}
is another positive function  (before recollimation) that depends only
on the MHD wind solution, while
\begin{eqnarray}
F(T)={kT(1+f_{\rm e})}{{m_{in} f_{\rm i} f_{\rm n}
<{\sigma_{in}v}>}}
\times {\Big( \frac{ \rho }{{\rho}_n } \Big)^2}
\end{eqnarray}
depends only on the local temperature and ionization state of the gas.

One can see that $F(T)$ always increases as a function of temperature:
either  linearly  $\propto T$  \fion\  when  ionization  is fixed  (by
photoionization or freeze-out), or  much more steeply when collisional
ionization of  H is efficient,  around $10^4$~K. These two  regimes of
$F(T)$ are illustrated in  Fig.~\ref{fig:plateau} in the case of local
ionization  equilibrium. The  function  $G(\chi)$ is  also plotted  in
Fig.~\ref{fig:plateau} for our solutions. It increases very rapidly at
the  base of  the flow  and then  stabilizes in  a plateau  beyond the
Alfv\'en point.

Now let us define a temperature $T_{\rm c}(\chi)$ such that:
\begin{eqnarray}\label{eq:Tc}
F(T_{\rm c})= G(\chi) &\Longleftrightarrow &\Gamma_{\rm{drag}}
= \Lambda_{\rm{adia}}.
\end{eqnarray}
Since $F$ increases  with temperature, it can be  readily seen that if
at a given point $T> T_{\rm c}(\chi)$, then $G(\chi)/F(T) < 1$ and the
gas will cool.  Conversely, if  $T<T_{\rm c}(\chi)$, the gas will heat
up.  Thus, the  fact that  $F(T)$ is  a rising  function  introduces a
feedback  that {\it  tries} to bring the temperature  near its  local
equilibrium value $T_{\rm c}(\chi)$, and $\Lambda_{\rm adia}$
near $\Gamma_{\rm drag}$.

However,  $T \simeq  T_{\rm c}(\chi)$  is not  necessarily  a possible
solution of Eq.~\ref{eq:amb_temp}: in that case the right-hand term is
close to zero (by definition of $T_{\rm c}$ in Equ.~\ref{eq:Tc}) hence
one must also have $d \ln T_{\rm c}/d \ln \chi \ll 1$ for consistency:
Only when $T_{\rm  c}(\chi)$ is flat can $T$  converge to $T_{\rm c}$,
and can  we have a temperature plateau  with $\Lambda_{\rm adia}\simeq
\Gamma_{\rm drag}$.

Indeed, $T_{\rm c}$ is, for our models, a flat function. We can understand
this by differentiating Eq.~\ref{eq:Tc}:
\begin{eqnarray}
\frac{d  \ln G}{d  \ln \chi}  = \left(\frac{d  \ln F}{d  \ln T}\right)_{T =
T_{\rm c}}  \left( \frac{d \ln
T_{\rm c}}{d \ln \chi} \right).
\end{eqnarray}
Hence $|d  \ln T_{\rm c}/d \ln \chi|  \ll 1$ is equivalent  to $|d \ln
G/d  \ln \chi| \ll  |d \ln  F/d \ln  T|$.  This  is fulfilled  for our
models: Below the Alfv\'en surface,  $G$ varies a lot, but collisional
H  ionization is  sufficiently  close to  ionization equilibrium  that
$F(T)$       still      rises       steeply       around      $10^4$~K
(cf. Fig.~\ref{fig:plateau}). For our numerical values of $G$, we have
$T_{\rm c}  \simeq 10^4$~K and thus $|d  \ln G/d \ln \chi|  \ll |d \ln
F/d \ln T|$.  Further out, where ionization is frozen  out, we have $d
\ln F/d \ln  T = 1$ (because $F\propto T$\fion) but  it turns out that
in this region $G$ is a slowly varying function of $\chi$, and thus we
still have $|d \ln G/d \ln \chi| \ll |d \ln F/d \ln T|$.

The  behavior  of  \fprot\  when  $G=F$ can  also  be  understood:  in
self-similar     disk      wind     models,     $G$      scales     as
$1/$($\varpi_0$\Macc).  Since $F(T) \propto  T$\fprot\ (when  \fe $\ll
1$)  and  \fprot\  is  a   high  power  of  $T$  close  to  ionization
equilibrium, most  of this  scaling is absorbed  by \fprot,  while $T$
remains around  $10^4$~K for  a wide range  of model  parameters. Note
also  that inclusion  of thermal  speeds (cf.  Draine  1980) increases
$<\sigma_{in}  v>$  by $\simeq$  10  compared  to  Safier (1993a)  and
decreases our  values of  \fprot\ by the  same amount, since  only the
product of these two terms enters in $F(T)$.

We conclude  that MHD winds heated  by ambipolar diffusion  have a hot
temperature plateau only when several conditions are met: (1) $G$ must
be such  that $F=G$ happens around  $10^4$~K, (2) the wind  must be in
ionization  equilibrium or  near it  in regions  where $G$  is  a fast
function of  $\chi$; (3)  once we have  ionization freezing,  $G$ must
vary  slowly. For  example,  in the  models  of Ruden  et al.  (1990),
ionization  is  quickly  frozen   while  $G  \propto  1/r$,  hence  no
temperature  plateau can be  established (one  has instead  $T \propto
1/r$).

\section{Predicted forbidden line emission and comparison with observations}
\label{sec:emission}

\begin{figure*}[t]
\begin{center}
  \resizebox{2.2\columnwidth}{!}{\includegraphics{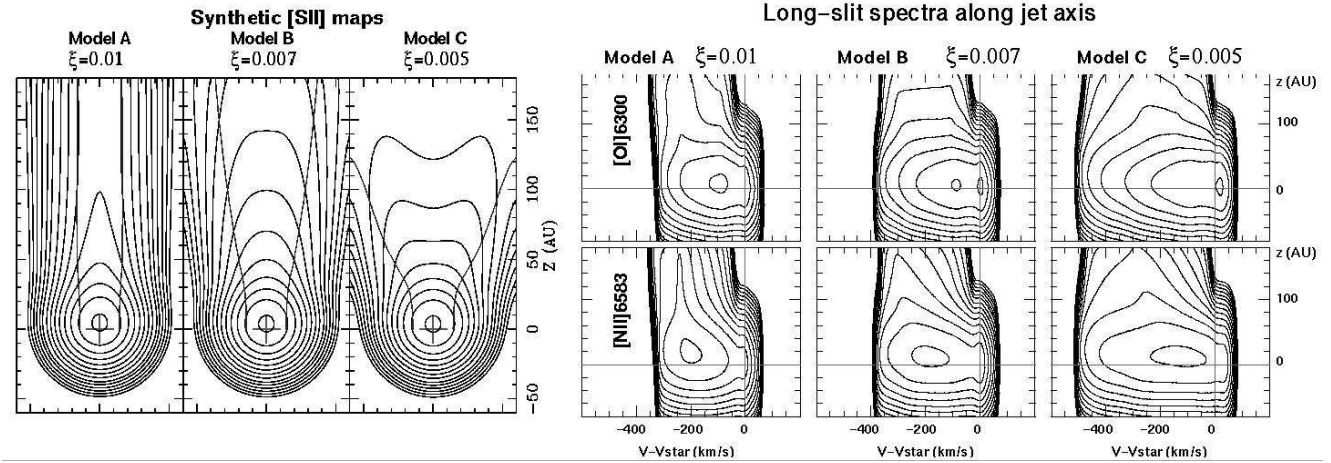}}
\vspace{-1cm}
\end{center}
\caption{
Emission maps  ({\bf left}) and long  slit spectra along  the jet axis
({\bf right})  for the  all models and  several forbidden  lines.  The
inclination  angle is  $i=60^\circ$ from  pole-on, and  mass accretion
rate is $\dot  M_{\rm acc}=10^{-6} M_{\odot} \rm{yr}^{-1}$.  Intensity
maps are convolved with a 28~AU resolution beam (0.2\arcsec in Taurus)
and  long-slit   spectra  are   convolved  by  $70   \,\rm{AU}  \times
10\,\rm{km/s}$, representative  of current ground-based spectroimaging
performances. Contours decrease by factors of 2.}
\label{fig:jet_tout}
\end{figure*}

We  compute  observational predictions  assuming  that  the disk  wind
extends from  0.07~AU (typical  disk corotation radius  for a  T Tauri
star)  to  1~AU.   Inside  corotation,  the disk  is  expected  to  be
truncated by  the stellar magnetosphere,  while beyond 1~AU,  the wind
should become molecular and  not contribute to forbidden line emission
(Safier 1993a).

Given the strong gradients in physical conditions present in the wind,
care  was taken  to convolve  synthetic  maps and  spectra to  typical
resolutions,  to allow  meaningful comparison  with  observations. The
successes and failures of our model predictions are outlined below.

\subsection{Wind morphology: Emission maps}

\begin{figure}[t]
  \resizebox{\columnwidth}{!}{\rotatebox{-90}{
      \includegraphics{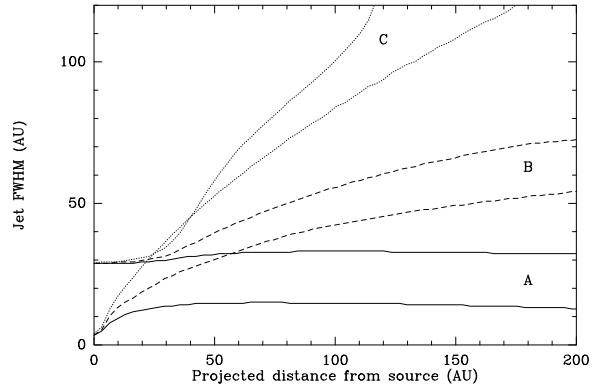}}}
  \caption{
Predicted jet transverse FWHM (obtained with gaussian fits) for models
A, B, C (with $\xi$  = 0.01,0.007,0.005) and two different beam sizes:
28 AU and 2.8 AU corresponding in Taurus to 0.2\arcsec and 0.02\arcsec
(VLT diffraction limit).  Note the  strong dependence on $\xi$ and the
bias introduced by the beam size.}
\label{fig:jet_fwhm}
\end{figure}

The  predicted maps  for our  models (Fig.~\ref{fig:jet_tout})
successfully reproduce two features present in observed microjets: (1)
an  unresolved peak slightly  shifted from  the stellar  position, (2)
jet-like emission appearing collimated within 200~AU of the star. Note
that the  MHD disk wind solutions  considered by Safier  1993a did not
collimate over such scales.

The shift of the unresolved peak  cannot be used as model test because
observed  shifts   are  severely  biased  by   accuracy  of  continuum
subtraction  (2\% errors  in  subtraction imply  $\sim4$~AU error  for
bright jets  \cite{garcia99}) and by possible obscuration  by a flared
disk.

The jet FWHM, in contrast,  provides an excellent quantitative test of
the  model:  It  is  almost  independent  of  the  inclination  angle,
accretion  rate  and  line,  but  is strongly  dependent  on  the  MHD
dynamical    solution   (Figs.~\ref{fig:jet_tout},\ref{fig:jet_fwhm}).
Note   that  it   is   also   severely  biased   by   the  beam   size
(Fig.~\ref{fig:jet_fwhm}), an illustration of the importance of proper
convolution before comparing to observations.

Predicted jet FWHM for models A  and B ($\xi = 0.007-0.01$) agree very
well with  observations of  T Tauri microjets  at the  same resolution
(Dougados et  al., 2000 and in  this volume). Models  with lower $\xi$
(e.g. model  C) are excluded.  Note that at sufficiently  high spatial
resolution,  disk  jets  should  appear  hollow,  hence  it  would  be
important to get very high angular resolution data.

\subsection{Wind kinematics: Line profiles}

\begin{figure}
\begin{center}
  \resizebox{\columnwidth}{!}{\includegraphics{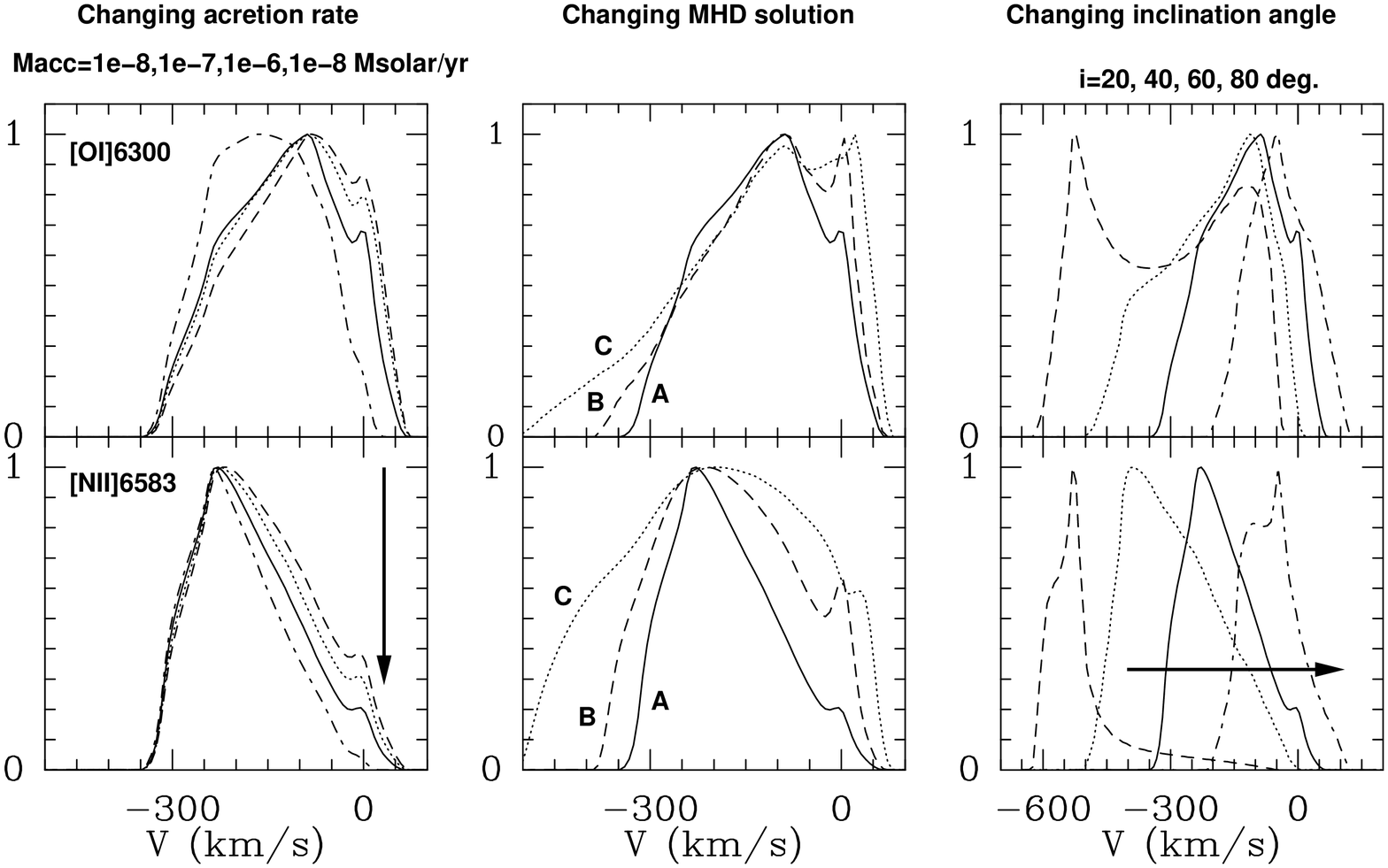}}
\vspace{-1cm}
\end{center}
  \caption{Influence of model parameters on line  profiles. {\bf Left:} 
effect  of accretion  rate.   The  arrow points  in  the direction  of
increasing   $\dot{M}_{\rm  acc}$.   {\bf   Middle:}  effect   of  MHD
solution. {\bf  Right: }effect of inclination angle.  The arrow points
in the direction of increasing $i$.}
\label{fig:profs}
\end{figure}

Typical  long-slit  spectra  are   presented  in  the right panels of
Fig.~\ref{fig:jet_tout}, and  integrated spectra for a  wider range of
parameters are  plotted in Fig.~\ref{fig:profs}.   The predictions are
successful in several respects:

- A  compact low-velocity component  (LVC) is  produced near  the star
(Fig.~\ref{fig:jet_tout}),  originating  at  the  base of  the  slowly
rotating outer disk  streamlines. The LVC is stronger  in \oi\ than in
\nii, and is stronger  for low accretion rates (Fig.~\ref{fig:profs}),
as observed (HEG95).

- An extended  high-velocity component (HVC) is  also present, tracing
the accelerated regions of the  wind. It is displaced further from the
stellar position than the LVC, and the displacement is larger in \nii\
than  in   \oi\  (Fig.~\ref{fig:jet_tout}),  as   observed  (Hirth  et
al. 1997).  The \nii\ line profile  peaks at the blue edge of the \oi\
line profile (Fig.~\ref{fig:profs}), as also observed (HEG95).

However, we fail to reproduce some observed features:

- The relative  intensity of the  LVC with respect  to the HVC  is too
small.   At $i  < 60^o$,  it  does not  appear anymore  as a  distinct
component in the \oi\ profile. This weakness of the LVC stems from the
low  temperature  and ionization  below  the  Alfv\'en surface,  where
low-velocity gas is located.

- Predicted centroid  and maximum  radial velocities of  \oi\ profiles
agree well with observations of the  DG Tau microjet, but are too high
compared  with typical  TTS  profiles (HEG95)  unless  most stars  are
observed  at   $i  \ge  80^o$   (see  Fig.~\ref{fig:profs}).  Terminal
velocities   in   cold   disk   wind   models   are   $\simeq   V_{\rm
kep}(\varpi_0)/\sqrt{\xi}$, so even our highest $\xi = 0.01$ (model A)
gives excessive speeds.

\subsection{Integrated line fluxes}

Several  trends  are  well  reproduced:  Predicted  integrated  fluxes
increase linearly with accretion rate, as observed (HEG95). This is an
interesting consequence of the weak  dependence of \nel\ in the jet on
\Macc\  (cf.  Fig.~1):  emissivity  ($\propto$  \nel  $n_H$)  is  then
proportional to  total density $n_H$,  instead of $n_H^2$ as  would be
the  case  for  a  constant  ionization  fraction.   We  also  find  a
correlation between integrated  fluxes in \oi\ and \sii,  with a slope
close to that observed (see Cabrit et al. 1990).

Quantitatively, however, integrated fluxes are systematically too weak
compared to observations of TTS,  {\it for the same range of accretion
rates}, i.e.   $10^{-8}-10^{-5}$ \msunyr  (HEG95).  We find  L(\oi) in
$L_\odot$ $\simeq$ \Macc\ in  \msunyr.  The discrepancy is typically a
factor 30  for a given \Macc.  In  the next section, we  show that the
flux  deficit stems  from insufficient  \nel, but  also  possibly from
insufficient total density $n_H$ in our models.

\subsection{Forbidden line ratios and total wind densities}

As noted by  Bacciotti \& Eisloeffel (1999), forbidden  line ratios of
\sii6716/6731, \nii/\oi, and \sii/\oi\  reflect directly the values of
\nel, \fe, and $T$. Hence, they test mainly the heating mechanism, not
the underlying dynamical solution.

Spatially resolved line ratios within 200~AU of the star are available
for four TTS microjets: HH~30, Th~28, DG~Tau, and RW~Aur (Bacciotti \&
Eisloeffel 1999;  Lavalley-Fouquet et al.  2000; Bacciotti et  al. and
Dougados et  al., this volume). Comparison with  our predictions, {\it
at the  same beam  resolution and distance  from the  star}, indicates
that our models have slightly too high temperatures, and beam-averaged
\nel\ and \fe\  that are too low  by up to a factor  30. Two important
remarks are in order:

(1) because   of  volume   effects  and   strong   density  gradients,
beam-averaged line  ratios differ greatly from local  line ratios, and
give very little weight to the inner densest parts of the wind.  Hence
it  is essential  to apply  proper convolution  to  perform meaningful
comparison with observed ratios.

(2) No ad-hoc heating  rate tuning is present in  our model. Predicted
line  ratios  would improve  if  we  include  extra mechanical  energy
deposition, e.g.  in shocks, as  illustrated in Dougados et  al. (this
volume), or with  a dissipation length prescription, as  done in Shang
et al. (this volume).

For the latter reason, the total  density $n_H = $ \nel/\fe\ is a more
useful quantity  that can be directly compared  with model predictions
independently  of the  ionization process  (although it  can  still be
affected by  shock compression and by beam  averaging).  For accretion
rates  of   $10^{-6}$  \msunyr,  total  densities   on  the  innermost
streamline for our  models are too low by a  factor 10-100 compared to
observations  of  the  above  four  bright  T~Tauri  microjets.  Shock
compression could explain  a factor of 10 (Hartigan  et al. 1994), but
it seems that  our models have intrinsically too  low density close to
the star.

\section{Concluding remarks}
\label{sec:discuss}

\begin{figure}[t]
\begin{center}
  \resizebox{\columnwidth}{!}{\includegraphics{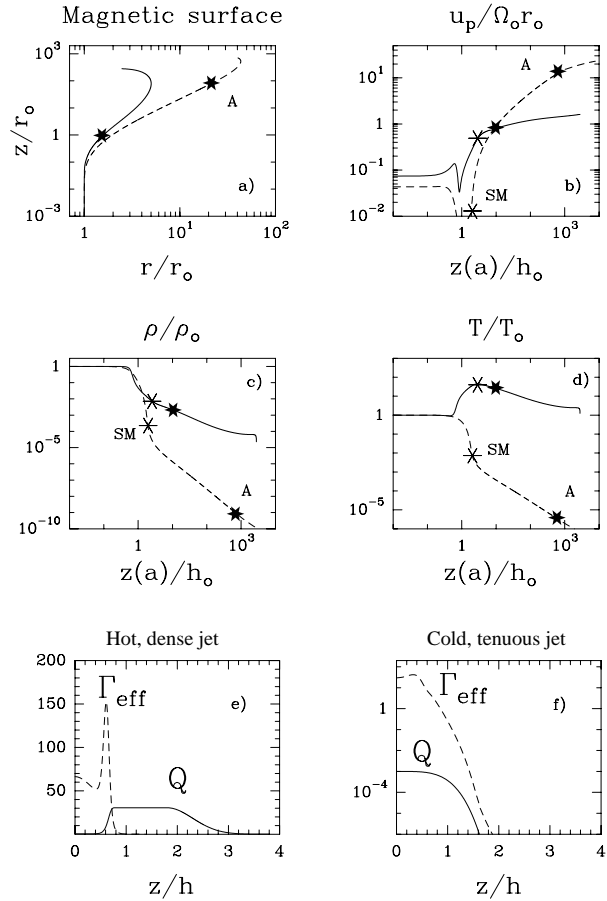}}
\vspace{-1cm}
\end{center}
\caption{{\bf Upper 4 panels:} New MHD disk wind solutions with large
heating at the disk surface (solid curves) have: more collimated streamlines
(top left), lower terminal velocity (top right), and higher density (middle
left) than a model with small heating (dashed curves). {\bf Bottom panels:}
Local energy dissipation rate (Joule and viscous; $\Gamma_{\rm eff}$), and
prescribed entropy injection $Q$ for the 2 models displayed above. See
Casse \& Ferreira (2000) for details. }
\label{fig:casse}
\end{figure}

We extended  the original  work of Safier  (1993a,b), and  compute the
thermal structure of a disk  wind jet, assuming current dissipation in
ion-neutral  collisions-- ambipolar  diffusion, as  the  major heating
source.  Improvements include: a)  detailed dynamical models where the
disk is self-consistently taken  into account; b) ionisation evolution
for  all relevant ``heavy  atoms''; c)  radiation cooling  by hydrogen
lines, recombination  and photoionisation heating;  d) correct $H-H^+$
momentum exchange  rates; and e)  more detailed dust  description.  We
still obtain warm  jets $10^4$~K but with ionisations  fractions 10 to
100 times smaller than Safier, due to larger $H-H^+$ momentum exchange
rates (including  the dominant  thermal velocity contribution)  and to
different MHD wind dynamics.

We have  presented a  complete set of  predictions for  forbidden line
emission,  where   we  have  stressed  the  crucial   effect  of  beam
convolution. The model reproduces  several observed trends: (1) images
show an  unresolved peak and  an extended high-velocity jet,  of width
compatible   with   observations.   (2)   Line   profiles  present   a
low-velocity  component  (LVC),  compact  and  near the  star,  and  a
high-velocity  component  (HVC),   tracing  the  jet.  The  systematic
differences between  \oi\ and \nii\ profiles are  reproduced. (3) Line
fluxes are  proportional to  $\dot{M}_{\rm acc}$, and  the \oi  - \sii
correlation slope is recovered.

Other line  ratios (\sii6716,6731, \nii/\oi) are  not well reproduced,
but we  stress that they  trace only the excitation  conditions, which
depend mainly on the heating mechanism, and do not test the underlying
dynamical model; a better fit  could be readily obtained with an extra
tunable heating rate, such as done  e.g. by Shang et al. (this volume)
or with shocks (Dougados et al., this volume).

Two  intrinsic  properties  of  our  dynamical models  are  not  fully
successful: Wind terminal velocities appear too high, and densities at
the wind base  appear too small, when compared  with current estimates
in bright microjets.   One possible improvement would be  to relax the
assumption  of a  cold  disk wind,  and  include heating  at the  disk
surface \cite{casse}.   MHD disk winds  with higher density  and lower
terminal     velocities     can      then     be     obtained     (see
Fig.~\ref{fig:casse}). The self-consistent  thermal structure of these
solutions remains to be investigated.

\acknowledgements 

%% When using the rmaacite package, the \bibitem command should be of
%% the format: 
%%
%%             \bibitem[AUTHOR<YEAR>]{KEY} 
%%
%% so that the \cite{KEY}, etc. commands will work properly. 
%% 
%% If you are doing the citations manually, then you can just use
%% `\bibitem{}' instead. This will give you a warning about
%% `multiply-defined labels' which you can safely ignore.
%% 

\end{document}